\documentclass[11pt,fleqn]{article}
\usepackage{graphicx}

\begin{document}

\title{
Asymptotic Pad\'e-Approximant Predictions for Renormalization-Group 
Functions of
Massive  $\phi^4$ Scalar Field Theory
}

\author{
{F. Chishtie, V. Elias\thanks{email: zohar@apmaths.uwo.ca}}\\
{ Department of Applied Mathematics,}
{ The University of Western Ontario,} \\
{ London, Ontario N6A 5B7, Canada. }
\and
T.G. Steele\thanks{email: steelet@sask.usask.ca} \\
{ Department of Physics and Engineering Physics and}\\
{ Saskatchewan Accelerator Laboratory}\\
{ University of Saskatchewan}\\
{ Saskatoon, Saskatchewan S7N 5C6, Canada.}
}
\maketitle

\begin{abstract}
      Within the context of massive $N$-component $\phi^4$ scalar field 
theory, we use asymptotic
Pad\'e-approximant methods to estimate from prior orders of perturbation 
theory the five-loop
contributions to the coupling-constant $\beta$-function $\beta_g$, the 
anomalous mass dimension $\gamma_m$, the
vacuum-energy $\beta$-function $\beta_v$, and the anomalous dimension 
$\gamma_{_2}$ of the scalar field propagator.
These estimates are then compared with explicit calculations of the 
five-loop contributions to $\beta_g$,
$\gamma_m$, $\beta_v$, and are seen to be respectively within 5\%, 18\%, 
and 27\% of their true values for $N$
between 1 and 5. We then extend asymptotic Pad\'e-approximant methods to 
predict the presently
unknown six-loop contributions to $\beta_g$, $\gamma_m$, and $\beta_v$. 
These predictions, as well as the six-loop
prediction for $\gamma_{_2}$, provide a test of asymptotic 
Pad\'e-approximant methods against future
calculations.
\end{abstract}

A substantial body of work \cite{pade1,pade2,pade3,pade4,pade5} already 
exists in which Pad\'e-approximant
methods are utilized to predict higher order corrections in quantum field 
theoretical calculations.
In the present letter, we apply such methods to the coupling-constant 
$\beta$-function $\beta_g(g)$, the
anomalous mass dimension $\gamma_m(g)$, the anomalous scalar-field 
propagator dimension $\gamma_{_2}(g)$, and the
vacuum-energy $\beta$-function $\beta_v(g)$ for a massive $\phi^4$ 
$N$-component scalar field theory based on the
Lagrangian\footnote{The final constant term in (\protect\ref{lagrangian}) 
is of relevance only for the calculation of
$\beta_v(g)$ \cite{kastening}, and has not been incorporated in 
calculations \cite{kleinart} of
$\beta_g(g)$, $\gamma_m(g)$, and $\gamma_{_2}(g)$.}
\begin{equation}
{\cal 
L}=\sum_{a=1}^N\left[\frac{1}{2}\partial_\mu\phi^a\partial_\mu\phi^a+\frac{1}{2}m^2\phi^a\phi^a
+\frac{16\pi^2}{4!}g\left(\phi^a\phi^a\right)^2\right]\quad 
\left(+\frac{m^4h}{(4\pi)^2g}\right)
\label{lagrangian}
\end{equation}
Specifically, we utilize the asymptotic Pad\'e algorithm described in refs. 
\cite{pade2} and \cite{pade3} to compare
predictions of five-loop contributions to $\beta_g(g)$, $\gamma_m(g)$, and 
$\beta_v(g)$ to their now-known \cite{kleinart,kastening2}
calculated values. These predictions are seen to be startlingly successful 
for $\beta_g(g)$, and for up to
five scalar field components, are generally within 20\% of the correct 
value for the five-loop term
in $\gamma_m(g)$ and $\beta_v(g)$.  We then adapt asymptotic Pad\'e methods 
to predict the presently unknown
six-loop contributions to 
$\gamma_2 (g)$, $\beta_g(g)$, $\gamma_m(g)$, and $\beta_v(g)$, with the hope
that these predictions may be tested against six-loop calculations in the 
not too distant future.

The asymptotic Pad\'e approximant methods we employ are directed toward 
perturbative
field-theoretical series of the form
\begin{equation}
F(x)=F_0(x)\left[1+\sum_{k=1}^\infty R_kx^k\right]
\label{basic_series}
\end{equation}
where $x$ typically denotes a field-theoretical coupling constant.
 Unknown coefficients $R_{_{N+M+1}}$ are estimated via $[N|M]$ Pad\'e 
approximants whose
Maclaurin expansions reproduce known coefficients $R_1$ -- $R_{_{N+M}}$:
\begin{eqnarray}
F(x)&\simeq& F_0(x)\frac{1+a_{_1} 
x+a_{_2}x^2+\ldots+a_{_N}x^N}{1+b_1x+b_2x^2+\ldots+b_{_M}x^M}
\nonumber\\
&=&F_0(x)\left[1+R_1x+R_2x^2+\ldots+R_{_{N+M}}x^{N+M}+R^{Pade}_{_{N+M+1}  
}x^{N+M+1}+{\cal O}\left(x^{N+M+2}\right)\right]\quad .
\label{series}
\end{eqnarray}
Within a quantum field-theoretical context in which coefficients $R_k$ are 
expected to diverge like
$k! C^kk^\gamma$ \cite{vainshtein}, it has been argued 
\cite{pade2,pade4,pade5} that the relative error in using an $[N|M]$ Pad\'e
approximant to estimate the coefficient $R_{_{N+M+1}}$ satisfies the 
asymptotic error formula
\begin{equation}
\delta_{_{N+M+1}}\equiv\frac{R^{Pade}_{_{N+M+1}}-R_{_{N+M+1}}}{R_{_{N+M+  
1}}}\simeq
-\frac{M!A^M}{\left(N+M+aM+b\right)^M}\quad ,
\label{asymp_error}
\end{equation}
where $\left\{A,~a,~b\right\}$ are constants to be determined.  Within the 
context of $[N|1]$ approximants
({\it i.e.} $M= 1$), the Pad\'e prediction for $R_{_{N+2}}$ is 
$R^2_{_{N+1}}/R_{_{N}}$ , and the error formula (\ref{asymp_error}) 
simplifies to
\begin{equation}
\delta_{_{N+2}}=\frac{R^2_{_{N+1}}-R_{_N}R_{_{N+2}}}{R_{_N}R_{_{N+2}}}=-  
\frac{A}{N+1+(a+b)}\quad .
\label{delta_N+2}
\end{equation}
Noting that $R_0\equiv 1$ in (\ref{basic_series}), one easily sees from 
(\ref{delta_N+2}) that \cite{pade2,pade3}
\begin{equation}
A=\frac{\delta_2\delta_3}{\delta_3-\delta_2}\quad ,\quad 
(a+b)=\frac{\delta_2-2\delta_3}{\delta_3-\delta_2}
\label{constants}
\end{equation}
We further note from (\ref{delta_N+2}) that the $[2|1]$ Pad\'e prediction 
for $R_4$ can be corrected by the asymptotic
error formula to yield the following result \cite{pade3}:
\begin{equation}
R_4=\frac{R_3^2/R_2}{1+\delta_4}=\frac{R_3^2\left[3+(a+b)\right]}{R_2\left[3-A+(a+b)\right]}
=\frac{R_3^2\left[R_2^3+R_1R_2R_3-2R_1^3R_3\right]}{R_2\left[2R_2^3-R_1^  
3R_3-R_1^2R_2^2\right]}
\label{R_4}
\end{equation}
The final expression in (\ref{R_4}) is obtained via (\ref{constants}), with 
$\delta_2= \left(R_1^2 - R_2\right)/R_2$,
$\delta_3 = \left(R^2_2/R_1 - R_3\right)/R_3$.

      We first test this asymptotic Pad\'e-approximant prediction (APAP) 
for $R_4$ against the
known values of the coefficients $R_1\ldots R_4$ in the coupling-constant 
$\beta$-function for the scalar field theory (\ref{lagrangian}):
\begin{equation}
\beta_g(g)=\frac{g^2(N+8)}{6}\left[1+\sum_{k=1}^\infty R_k g^k\right]\quad 
.
\label{beta_g}
\end{equation}
The coefficients $R_1$, $R_2$, $R_3$ {\em and} $R_4$ have been calculated 
explicitly \cite{kleinart}, and are tabulated in Table
\ref{beta_g_table}.  The APAP prediction for $R_4$, as obtained via 
(\ref{R_4}) from prior coefficients ($R_1\ldots R_3$) is also tabulated
as $R_4^{APAP}$ in
Table \ref{beta_g_table}.  The relative error
\begin{equation}
E\equiv\left\vert\frac{R_4-R_4^{APAP}}{R_4}\right\vert
\label{error}
\end{equation}
is seen to be astonishingly small (0.2\%) for single-component ($N=1$) 
scalar field theory, and remains less than
5\% for $N\leq 5$.\footnote{The coefficients $R_4$ for $\beta_g$ have also 
been estimated quite accurately by a somewhat different procedure
in \cite{pade5} involving explicit use of the $N^4$ dependence of $R_4$, as 
well as an overall fit of $R_4$'s $N$-dependence within the context of a 
simplified
version of the error formula (\ref{R_4}).}

      We have repeated the above procedure for coefficients $R_1$--$R_4$, 
as extracted from  ref. \cite{kleinart}, in the
anomalous mass dimension $\gamma_m(g)$:
\begin{equation}
\gamma_m(g)=\frac{g(N+2)}{6}\left[1+\sum_{k=1}^\infty R_k g^k\right]\quad .
\label{gamma_m}
\end{equation}
These values are tabulated in Table \ref{gamma_m_table} for $N = 
\{1,2,3,4,5\}$.  The APAP prediction for $R_4$, as
obtained via (\ref{R_4}) from ${R_1, R_2, R_3}$, is within 10.4\% of the 
true value for a single component
scalar field ($N = 1$), and becomes progressively less accurate as the 
number of scalar field
components increases.  Nevertheless, even the five-component APAP estimate 
of $R_4$ differs from
the true value by only 18.3\%, remarkable accuracy for a five-loop estimate 
that is obtained {\em
entirely} from lower-order perturbative contributions.

      We suspect the somewhat diminished accuracy in the estimate of the 
five-loop
contribution of $\gamma_m$, as compared to that for $\beta_g$, may be a 
consequence of the value for $R_1$
remaining static at $-5/6$ while $R_2$, $R_3$, and $R_4$ are all seen to 
increase in magnitude with the
number of scalar-field components $N$.  Indeed, a similar situation arises 
for the vacuum-energy
$\beta$-function $\beta_v(g)$, as defined and calculated in 
\cite{kastening} and \cite{kastening2} for the $N$-component massive 
scalar-field theory
(\ref{lagrangian}):
\begin{eqnarray}
\beta_v(g)&=&\frac{Ng}{4}\biggl[1+\frac{N+2}{24}g^2-0.032639868(N+2)(N+8  
)g^3\biggr.
\nonumber \\
& &\qquad 
\biggl.+4(0.5781728935+0.4184309753N+0.06839310856N^2+0.001860422132N^3)  
g^4+\ldots\biggr]
\nonumber\\
&\equiv &\frac{Ng}{4}\left[1+\sum_{k=1}^\infty R_kg^k\right]
\label{beta_v}
\end{eqnarray}
In the above expression, $R_1$ is clearly zero for all $N$.
Consequently, APAP estimates for $R_4$ obtained via (\ref{R_4}) are simply 
$R_3^2/(2R_2)$.  Surprisingly, such an
estimate does substantially better than $R_3^2/R_2$, the naive estimate of 
$R_4$ from the $[2|1]$-approximant, as
is evident from Table \ref{beta_v_table}.  We note that $R_4^{APAP}$ 
actually improves in accuracy with increasing $N$; the relative
error in $R_4^{APAP}$ decreases from 27\% to 18\% as $N$ increases from 1 
to 5.

The six-loop contribution to the scalar-field anomalous dimension 
$\gamma_{_2}$ has not, to our
knowledge, been calculated.  Only contributions up to five-loop order are 
listed in ref. \cite{kleinart}.
These results, when expressed in the form
\begin{equation}
\gamma_{_2}(g)=\frac{g^2(N+2)}{36}\left[1+\sum_{k=1}^\infty R_k g^k\right]
\label{gamma_2}
\end{equation}
are tabulated in Table \ref{gamma_2_table}.  The algorithm (\ref{R_4}) is 
used to predict $R_4$, the presently unknown
six-loop contribution.  A genuine test of the asymptotic error formulae
 (\ref{asymp_error}) would be to compare the
final column of Table \ref{gamma_2_table} with the results of a six-loop 
calculation.

      It is perhaps of even greater value to see if the success already 
evident in APAP
predictions of five-loop terms for $\beta_g(g)$, $\gamma_m(g)$, and $\beta_v(g)$ carry over to the next order of
perturbation theory, by extending the APAP procedure based on the error 
formula (\ref{asymp_error}) to estimate
six-loop order contributions ($R_5$) from lower-order terms in perturbation 
theory. We note from
(\ref{delta_N+2}) that
\begin{equation}
R_5=\frac{R_4^2}{R_3\left(1+\delta_5\right)}=\frac{R_4^2\left[4+(a+b)\right]}{R_3\left[4-A+(a+b)\right]}\quad .
\label{R_5_cons}
\end{equation}
We seek to improve upon (\ref{constants}) by incorporating knowledge of 
$R_4$, hence of $\delta_4$, into
$A$ and $(a+b)$.  Since this applies the  asymptotic error formula at a 
larger value of $N$, the results of the constants in
(\ref{asymp_error}) should be more accurate.
The equations
\begin{equation}
\delta_3=-\frac{A}{2+(a+b)}\quad ,\quad \delta_4=-\frac{A}{3+(a+b)}
\label{delta_5_eqns}
\end{equation}
have solutions
\begin{equation}
A=\frac{\delta_3\delta_4}{\delta_4-\delta_3}\quad ;\quad 
(a+b)=\frac{2\delta_3-3\delta_4}{\delta_4-\delta_3}
\end{equation}
Noting from (\ref{delta_N+2}) that
$\delta_4 =\left(R_3^2 - R_2R_4\right)/\left(R_2R_4\right)$, we obtain the 
asymptotic error formula prediction
\begin{equation}
R_5=\frac{R_4^2\left[R_1R_3^3-2R_4R_2^3+R_1R_2R_3R_4\right]}{R_3\left[2R  
_1R_3^3-R_4R_2^3-R_2^2R_3^2\right]}
\label{R_5}
\end{equation}
The final column of Table \ref{beta_g_table} lists this APAP prediction of 
the six-loop term $R_5$ for $N= \{1,2,3,4,5\}$
In view of the accuracy already noted in the APAP prediction of the 
five-loop contribution to
$\beta_g(g)$, a comparison of this column's entries to an explicit 
calculation of the six-loop
contributions to the $\beta$-function would be of evident value.

      For completeness, we have also added to Tables \ref{gamma_m_table} 
and \ref{beta_v_table} predictions of the six-loop
contributions to $\gamma_m(g)$ and $\beta_v(g)$, as obtained from 
(\ref{R_5}).  Kastening has already noted \cite{kastening2} that the
six-loop contribution to $\beta_v(g)$ is perhaps the easiest six-loop 
quantity in scalar field theory to
calculate.  Perhaps a direct comparison of such a calculation to the final 
column of Table \ref{beta_v_table} will
be possible in the near future.

      We are grateful for support from the Natural Sciences and Engineering 
Research Council
of Canada.

\begin{table}[htb]
\centering
\begin{tabular}{||c|c|c|c|c|c|c|c||}\hline\hline
$N$ & $R_1$ & $R_2$ & $R_3$ & $R_4$ & $R_4^{APAP}$ & $E$ & $R_5^{APAP}$ 
  \\ \hline\hline
1 & $-\frac{17}{9}$ & 10.84989 & -90.53526 & 949.5228 & 947.8 & 0.002 & 
-11744.2 \\ \hline
2 & -2 &  11.98433 & -105.1544 & 1153.352 & 1165.0 & 0.010 & -14814.4 \\ 
\hline
3 & $-\frac{23}{11}$ & 12.99584 & -119.3579 & 1363.260 & 1394.1 & 0.023 & 
-18116.2 \\ \hline
4 & $-\frac{13}{6}$ & 13.91515 & -133.2477 & 1579.416 & 1633.9 & 0.035 & 
-21659.1 \\ \hline
5 & $-\frac{29}{13}$ & 14.76355 & -146.8943 & 1801.925 & 1883.4 & 0.045 & 
-25451.9 \\ \hline \hline
\end{tabular}
\caption{
The coefficients $R_1$--$R_4$ in the coupling-constant $\beta$-function 
$\beta_g(g)$, as defined
in (\protect\ref{beta_g}) and as calculated in ref. \protect\cite{kleinart} 
for an $N$-component
massive $\phi^4$ scalar field theory.  The quantity $R_4^{APAP}$ is the 
asymptotic Pad\'e-approximant
estimate of the five-loop contribution $R_4$, and $E$ is the relative error 
defined in (\ref{error})
between this estimate and $R_4$'s true value.  The quantity $R_5^{APAP}$ is 
the asymptotic Pad\'e estimate
of the six-loop contribution to $\beta_g(g)$.
}
\label{beta_g_table}
\end{table}

\begin{table}[htb]
\centering
\begin{tabular}{||c|c|c|c|c|c|c|c||}\hline\hline
$N$ & $R_1$ & $R_2$ & $R_3$ & $R_4$ & $R_4^{APAP}$ & $E$ & $R_5^{APAP}$ 
  \\ \hline\hline
1 &$-\frac{5}{6}$ & $\frac{7}{2}$ & -19.956305 & 150.7555 & 135.07 & 0.104 
& -1478 \\ \hline
2 &$-\frac{5}{6}$ & $\frac{47}{12}$ & -23.90300 & 191.8840 & 168.37 & 0.123 
& -2055 \\ \hline
3 &$-\frac{5}{6}$ & $\frac{13}{3}$ & -27.98248 & 236.9432 & 203.06 & 0.143 
& -2882 \\ \hline
4 &$-\frac{5}{6}$ & $\frac{19}{4}$ & -32.19475 & 285.9396 & 239.18 & 0.164 
& -4738 \\ \hline
5 &$-\frac{5}{6}$ & $\frac{31}{6}$ & -36.53981 & 338.8796 & 276.80 & 0.183 
& +1446\\ \hline \hline
\end{tabular}
\caption{
The coefficients $R_1$--$R_4$ in the anomalous mass dimension 
$\gamma_m(g)$, as defined
in (\protect\ref{gamma_m}) and as calculated in  \protect\cite{kleinart}.
  The quantity $R_4^{APAP}$ is the asymptotic Pad\'e-approximant
estimate of the five-loop term $R_4$,  $E$ is the relative error in this 
five-loop term  [eq. (\ref{error})], and
$R_5^{APAP}$ is the asymptotic Pad\'e-approximant  estimate
of the six-loop contribution to $\gamma_m(g)$.
}
\label{gamma_m_table}
\end{table}

\begin{table}[htb]
\centering
\begin{tabular}{||c|c|c|c|c|c|c|c||}\hline\hline
$N$ & $R_1$ & $R_2$ & $R_3$ & $R_4$ & $R_4^{APAP}$ & $E$ & $R_5^{APAP}$ 
  \\ \hline\hline
1 & 0 &$\frac{1}{8}$ & -0.8812764 & 4.267430 & 3.1066 & 0.272 & -16.83 \\ 
\hline
2 &0& $\frac{1}{6}$ & -1.305595 & 6.813963 & 5.1137 & 0.250 & -28.43 \\ 
\hline
3 &0&$\frac{5}{24}$ & -1.795193 & 9.996941 & 7.7345 & 0.226 & -43.71 \\ 
\hline
4 &0& $\frac{1}{4}$ & -2.3500705 & 13.861014 & 11.0457 & 0.203 & -63.04 \\ 
\hline
5 &0& $\frac{7}{24}$ & -2.9702280 & 18.450833 & 15.1239 & 0.180 & -86.85 \\ 
\hline \hline
\end{tabular}
\caption{
The coefficients $R_1$--$R_4$ in the vacuum-energy $\beta$-function 
$\beta_v(g)$, as defined
in (\protect\ref{beta_v}).
Values of $R_1$, $R_2$, and $R_3$ are extracted from ref. \cite{kastening}; 
the five-loop contribution
$R_4$ is extracted from ref. \cite{kastening2}.
  The quantity $R_4^{APAP}$ is the asymptotic Pad\'e-approximant
estimate of the five-loop term $R_4$,  $E$ is the relative error in this 
five-loop term  [eq. (\ref{error})], and
$R_5^{APAP}$ is the asymptotic Pad\'e-approximant  estimate
of the six-loop contribution to $\beta_v(g)$.
}
\label{beta_v_table}
\end{table}

\begin{table}[htb]
\centering
\begin{tabular}{||c|c|c|c|c||}\hline\hline
$N$ & $R_1$ & $R_2$ & $R_3$ &  $R_4^{APAP}$    \\ \hline\hline
1 & $-\frac{3}{4}$ & $\frac{65}{16}$ & -23.10702 & 134.7 \\ \hline
2 & $-\frac{5}{6}$ & $\frac{55}{12}$ & -28.33255 & 186.1\\ \hline
3 & $-\frac{11}{12}$ & $\frac{725}{144}$ & -33.762556 & 252.0 \\ \hline
4 & -1 & $\frac{65}{12}$ & -39.37539 & 337.2 \\ \hline
5 & $-\frac{13}{12}$ & $\frac{275}{48}$ & -45.149425 & 448.9 \\ \hline 
\hline
\end{tabular}
\caption{
The coefficients $R_1$, $R_2$, $R_3$ in the  in the anomalous dimension 
$\gamma_{_2}$ of the scalar field
propagator, as defined
in (\protect\ref{gamma_2}) and extracted from ref. \cite{kleinart}.
The quantity $R_4^{APAP}$ is the asymptotic Pad\'e-approximant
estimate of the six-loop term.
}
\label{gamma_2_table}
\end{table}


\begin{thebibliography}{99}

\bibitem{pade1} J.-F. Yang, Comm. Theor. Phys. {\bf 22} (1994) 207; 
J.Ellis, E. Gardi, M. Karliner, M.A. Samuel, Phys. Lett. {\bf 366} (1996) 
268 and
Phys. Rev {\bf D54} (1996) 6986; I.T. Drummond, R. R. Horgan, P. V.
Landshoff, and A. Rebhan, Nucl. Phys. {\bf B524} (1998) 579.

\bibitem{pade2} J. Ellis, I. Jack, D.R.T. Jones, M. Karliner, M.A. Samuel, 
Phys. Rev. {\bf D57} (1998) 2665.

\bibitem{pade3} V. Elias, T.G. Steele, F. Chishtie, R. Migneron, K. 
Sprague, Phys. Rev. D (to appear: hep-ph/9806324).

\bibitem{pade4} M.A. Samuel, J. Ellis, M. Karliner, Phys. Rev. Lett. {\bf 
74} (1995) 4380.

\bibitem{pade5}  M.A. Samuel, J. Ellis, M. Karliner, Phys. Lett. {\bf B400} 
(1997) 176.

\bibitem{kastening} B. Kastening, Phys. Rev. {\bf D54} (1996) 3965.

\bibitem{kleinart} H. Kleinert, J. Neu, V. Schulte-Frohlinde, K.G. 
Chetyrkin, S.A. Larin, Phys. Lett. {\bf B272} (1991) 39 and {\bf B319} 
(1993) 545 (erratum).

\bibitem{kastening2} B. Kastening, Phys. Rev. {\bf D57} (1998) 3567; see also S. A. Larin, M..
Moenningmann, M. Stroesser, and V. Dohm, cond-mat/9805028.

\bibitem{vainshtein} A.I. Vainshtein, V.I. Zakharov, Phys. Rev. Lett. {\bf 
73} (1994) 1207.

\end{thebibliography}
\end{document}